\documentclass[runningheads]{llncs}
\usepackage{graphicx}
\usepackage{amsmath,amssymb,amsfonts}
\usepackage{algorithmic}
\usepackage{textcomp}
\usepackage[ruled,vlined]{algorithm2e}
\usepackage[dvipsnames]{xcolor}
\usepackage{booktabs}
\usepackage{hyperref}

\newcommand{\Sref}[1]{\S\ref{#1}}

\newcommand{\Fref}[1]{Figure~\ref{#1}}

\newcommand{\Tref}[1]{Table~\ref{#1}}

\makeatletter
\def\@fnsymbol#1{\ensuremath{\ifcase#1\or *\or \dagger\or \ddagger\or
   \mathsection\or \mathparagraph\or \|\or **\or \dagger\dagger
   \or \ddagger\ddagger \else\@ctrerr\fi}}
\makeatother

\begin{document}

\title{A Computational Analysis of Polarization on Indian and Pakistani Social Media}

\titlerunning{Polarization in India and Pakistan}
%
\author{Aman Tyagi\orcidID{0000-0002-6654-0670}\thanks{Equal Contribution} \and
Anjalie Field\orcidID{0000-0002-6955-746X}\textsuperscript{*} \and
Priyank Lathwal\orcidID{0000-0003-3883-3641} \and Yulia Tsvetkov\orcidID{0000-0002-4634-7128} \and Kathleen M. Carley\orcidID{0000-0002-6356-0238}}
\authorrunning{Tyagi et al.}
%
\institute{Carnegie Mellon University, Pittsburgh PA 15213, USA\\
\email{\{amant,anjalief,plathwal\}@andrew.cmu.edu, \{ytsvetko,kathleen.carley\}@cs.cmu.edu}}

\maketitle

\begin{abstract}
Between February 14, 2019 and March 4, 2019, a terrorist attack in Pulwama, Kashmir followed by retaliatory airstrikes led to rising tensions between India and Pakistan, two nuclear-armed countries. In this work, we examine polarizing messaging on Twitter during these events, particularly focusing on the positions of Indian and Pakistani politicians. We use a label propagation technique focused on hashtag co-occurrences to find polarizing tweets and users. Our analysis reveals that politicians in the ruling political party in India (BJP) used polarized hashtags and called for escalation of conflict more so than politicians from other parties. Our work offers the first analysis of how escalating tensions between India and Pakistan manifest on Twitter and provides a framework for studying polarizing messages.
\end{abstract}

\section{Introduction}
While social media platforms foster open communication and have the potential to offer more democratic information systems, they have simultaneously facilitated divisions in society by allowing the spread of polarizing and incendiary content \cite{demszky-etal-2019-analyzing,stewart2017drawing}. 
Polarizing content can be beneficial by encouraging pride and solidarity, but it has also become a social cyber-security concern: foreign and domestic actors may employ polarizing social media content to sow divisions in a country, to demean other nations, or to promote political agendas \cite{arif2018acting,bradshaw2018challenging,carley2018social,king2017chinese}.

Using automated methods to analyze social media offers a way to understand the type of content users are exposed to, the positions taken by various users, and the agendas pursued through coordinated messaging across entire platforms. Understanding the dynamics of this information landscape has become critical, because social media can strongly influence public opinion \cite{bradshaw2018challenging}. However, prior computational social science research on polarization has focused primarily on U.S. politics, and much attention has focused on the influence of Russian or Chinese state actors \cite{arif2018acting,badawy2018analyzing,demszky-etal-2019-analyzing,golovchenko2018state,king2017chinese,le2019postmortem,starbird2019disinformation,stewart2017drawing}.
In contrast, we focus on polarizing social media content in India and Pakistan and how it can contribute to rising tensions between these two nations. Specifically, we examine communication patterns on Twitter following the terrorist attack in the Pulwama district, Jammu and Kashmir, India, on February 14, 2019.

We primarily investigate: \textit{to what extent did entities on social media advocate for or against escalating tensions?}. India and Pakistan are both nuclear-armed countries and have a decades-long history involving multiple armed conflicts. The Pulwama attack in 2019 was followed by an escalation of tensions between these two nations that nearly approached full-fledged war \cite{feyyaz2019contextualizing,palakodety2019kashmir,pandya2019future}. Moreover, the relationship between these countries is an important agenda for political parties in both India and Pakistan. India has two primary political parties: the Indian National Congress (INC), which was dominant in the early 21\textsuperscript{st} century, and the Bharatiya Janata Party (BJP), which rose to prominence on a populist and nationalist platform in 2014 and has been in power since \cite{McDonnell2019BJP}. 
Given this context, we first examine the tweets and communication patterns of general users in order to understand how polarizing the attack was and to what extent users with different viewpoints may have interacted with each other. We then examine the social media messaging of political party members and how it changed over the sequence of events in order to uncover possible political agendas.

Our core methodology uses a network-based label propagation algorithm to quantify the polarity of hashtags along specified dimensions: Pro-India vs.~Pro-Pakistan and Pro-Aggression vs.~Pro-Peace. We then aggregate the hashtag-level scores into tweet-level and user-level scores, e.g. the polarity of a given user on a given day. Unlike methodology that assumes users' opinions do not change \cite{darius2019twitter,darwish2019quantifying,Weber2013}, focuses on binary stances \cite{Holthoefer2015}, or requires in-language annotations and feature-crafting \cite{magdy2016isisisnotislam}, our methodology allows us to analyze degrees of polarization in a multilingual corpus and how they change over time.

We begin by providing an overview of the events between February 14 and March 1, 2019 (\Sref{sec:events}). Next, we describe the Twitter data collection (\Sref{sec:data}) and discuss methods (\Sref{sec:method}) and evaluation (\Sref{sec:eval}). Our results (\Sref{sec:results}) suggest that more members of the BJP propagated a narrative of escalation than members of other political parties. This finding supports anecdotes reported by journalists \cite{WashP} about these events. Through this research, we develop (1) the first analysis of escalating tensions between India and Pakistan on Twitter, (2) a data-driven investigation of social media messaging following the 2019 Pulwama attack, and (3) a novel and general methodology to examine polarization on multilingual social media.

\section{Timeline of Events}
\label{sec:events}
 
We briefly provide background on relevant events, relying primarily on third party newspapers unaffiliated with either nation (The New York Times and BBC News) and noting where official accounts differ.

\textbf{Feb. 14, 2019} A 22-year old native of Pulwama 
carried out a suicide attack against a convoy carrying approximately 2,500 security personnel in the Pulwama district in Kashmir, India. The attack resulted in the death of more than 40 Indian soldiers. Jaish-e-Mohammad (JeM) a militant group based in Pakistan (the group is formally banned in Pakistan) claimed responsibility~\cite{bbcbackground,NYTConvoy}.

\textbf{Feb. 14-26, 2019} The Indian government responded to the attack with threats of retaliation against Pakistan, even though Pakistani officials denied any role~\cite{bbcThreats}. Diplomatic ties deteriorated, e.g., India revoked Pakistan's most favored nation status, which had provided trade advantages. 
Pakistan threatened to retaliate if India pursued military action \cite{bbcimrankhan}.

\textbf{Feb. 26, 2019} The Indian Air Force (IAF) conducted a retaliatory airstrike against a JeM training camp inside Pakistan, which the Indian government termed ``non-military, preemptive''~\cite{balakot}. According to Indian government officials, the JeM camp targeted by this airstrike was located 70km inside the Line of Control (LoC) -- the military line dividing the Indian and Pakistani controlled parts of Jammu and Kashmir. Indian officials reported that the airstrike was ``100 percent successful'', went on ``exactly as planned'', and killed over 200 terrorists~\cite{ndtvAirStrikes,balakot}. In contrast, Pakistani officials reported that the target of the attacks was located only 5--6km inside the LoC, that the Pakistani air force turned back the Indian fighters, and that the attacks landed in an empty area~\cite{dawn_indian_aircraft,tribune_report_strike}.

\textbf{Feb. 27, 2019} The Pakistan Air Force (PAF) carried out retaliatory airstrikes along the LoC. Indian and Pakistani officials presented different details of the strikes, but both emphasized de-escalation: a Pakistani official reported that the PAF intentionally targeted open spaces, to demonstrate Pakistan's capabilities without inviting escalation, while an Indian official reported no deaths or civilian casualties~\cite{cnnPAFStrike2,cnnPAFStrike1,econTimesPAF}. However, in  aerial combat following the strikes, an IAF pilot was captured by the Pakistani Army~\cite{bbcPilot,IAF}. 

\textbf{Mar. 1, 2019} Pakistan returned the IAF pilot to India on March 1 in what Pakistani Prime Minster Imran Khan called ``a gesture of peace''~\cite{bbcPilotRet}.

\section{Data}
\label{sec:data}
We collected tweets related to these events by first identifying a set of relevant hashtags.
Our hashtag set is based on hashtags related to \#pulwama found on \url{best-hashtags.com}.\footnote{\url{best-hashtags.com} uses an algorithm to provide popular hashtags that are similar to the provided seed (\#pulwama). Since our analysis, the website has stopped reporting Twitter hashtags.} We modified the hashtag set to ensure that it included both hashtags more likely to be used by Pro-India users (e.g., IndiaWantsRevenge) and hashtags more likely to be used by Pro-Pakistan users (e.g., PakistanZindabad). We then collected all tweets using these terms, either as words or as hashtags during the events.\footnote{We provide further details, including the full list of keywords, data statistics, network densities and evidence that our data set is comprehensive in our project repository: \url{https://github.com/amantyag/india_pakistan_polarization}.}

Our final data set contains 2.5M unique tweets (including retweets) from 567K users that use 67K unique hashtags. All tweets occurred between February 14\textsuperscript{th} and March 4\textsuperscript{th}. The data contains a mix of languages including English, Urdu, and Hindi, and many users use multiple languages in the same tweet. While some tweets express neutral opinions, others contain incendiary language, such as: \textit{``@PMOIndia @PMOIndia @narendramodi We r eagerly waiting for ur action of revenge...\#PulwamaRevenge \#IndiaWantsRevenge
\#PulwamaAttack''} and \textit{``I feel time has come to give all support to \#Balochistan activist. Let us \#bleed Pakistan from all fronts. \#NeverForget @PMOIndia @narendramodi \#IndiaWantsRevenge''}.


\section{Methodology}
\label{sec:method}
We develop a method to assign a polarity score to an aggregate group of tweets, and we analyze how polarities change over time for different groups of users. For instance, given pole $A$ (e.g., Pro-Pakistan) and pole $B$ (e.g., Pro-India), we aggregate all tweets by a given user and assign the user a polarity score between [a, b], where a score close to $a$ indicates the user more likely supports $A$ and a score close to $b$ indicates the user more likely supports B. We could also aggregate only tweets by the user on one day and determine the user's Pro-A/Pro-B polarity on that day.

In the absence of annotated data, we use a weakly supervised approach. First, for pole $A$, we hand-select a small seed set of tokens $S_A$ that are strongly associated with $A$, and we equivalently hand-select $S_B$. We assign each $s \in S_A$ a polarity score of $a$, and we assign each $s \in S_B$ a polarity score of $b$. Then, we use $S_A$ and $S_B$ to infer polarity scores over a larger lexicon of words or hashtags $\mathcal{V}$, where each $w \in \mathcal{V}$ is assigned a score in $[a, b]$. Finally, we estimate the polarity of an aggregated set of tweets by averaging the inferred polarity scores for all $w \in \mathcal{V}$ used in those tweets.

In order to propagate the hand-annotated labels in $S_A$ and $S_B$ to the larger lexicon $\mathcal{V}$, we use 3 variants of graph-based label propagation. In each variant, we construct a graph $G$, whose nodes consist of $w \in \mathcal{V}$ and whose edges and edge weights are defined based on similarity metrics between members of $\mathcal{V}$. We describe each variant in detail below.

\paragraph{Network-based Hashtag Propagation}
In the first variant, we define $\mathcal{V}$ to be the set of all hashtags used in our data set. Then, we construct $G$ as a hashtag*hashtag co-occurrence network. Each node in $G$ corresponds to a hashtag. Edges occur between hashtags that co-occur in the same tweet, and edge weights are proportional to how frequently the hashtags co-occur. Then, we use the label propagation algorithm detailed in Algorithm \ref{algo:label_prop} to infer polarity scores for $w \in \mathcal{V}$ from $S_A$ and $S_B$, where $a = -1$ and $b = 1$. The algorithm uses a greedy approach to assign labels to each node in \emph{G}. If all nodes connected to a node \emph{n} have been labeled, then node \emph{n} is assigned a weighted average of all the adjacent nodes. This step is repeated until the maximum possible number of nodes are labeled. A low value of $\gamma$ would label nodes neighboring unlabeled nodes, a high value would only label nodes neighboring unlabeled nodes after multiple iterations of the outer loop.

Our algorithm is similar to methods used to infer user-level polarities, in which a small seed of users is hand-annotated and a graph-based algorithm propagates labels to other users by assuming that users who retweet each other share the same views \cite{darwish2019quantifying,Weber2013}. For example, \cite{kiranMorales} quantify polarity based on a graph structure by assuming that the controversial topics induce clusters of discussions, commonly referred to as “echo-chambers”. However, we conduct propagation at a hashtag level, by assuming that hashtags that frequently occur in the same tweets indicate similar polarities. Also, our approach does not assume homophily in retweet network nor that user polarities are constant over time. Graph-based approaches have also been used to examine sentiment or for mixed tweet/hashtag/user-level analyses \cite{Coletto2016,Pollacci2017}.

\paragraph{Network-based Word Propagation}
The second variant is similar to the first; however, instead of restricting $\mathcal{V}$ to be the set of hashtags in the corpus, we define $\mathcal{V}$ to be the set of all tokens, including words and hashtags. We then construct G as a token*token co-occurrence network, and as above, we infer labels using Algorithm \ref{algo:label_prop} and obtain token-level polarity scores in the range [-1, 1]. Expanding $\mathcal{V}$ to all tokens instead of just hashtags allows our algorithm to incorporate more information, but also risks introducing noise, as we do not attempt to process nuances in language like negation.

\paragraph{Embedding-based Word Propagation (SentProp)}
In the third variant, we define $\mathcal{V}$ to be the set of all tokens, as in the Network-based Word Propagation approach. Then, we train GloVe embeddings~\cite{pennington2014glove} over our entire corpus (limiting vocabulary size to 50K). We then use SentProp~\cite{hamilton2016inducing}, a method for inferring domain-specific lexicons to infer labels over $\mathcal{V}$. In this method, as before, we construct a graph $G$ where each $w \in \mathcal{V}$ is a node. However, rather than relying on raw co-occurrence scores, SentProp uses embedding similarity metrics to define edge weights and a random-walk method to propagate labels. We implement SentProp using the SocialSent package \cite{hamilton2016inducing}, where $a = 0$ and $b = 1$.

\begin{algorithm}[ht]
\KwIn{Graph \emph{G} with nodes \emph{n} and edges \emph{e} with \emph{$e_{ij}$} as the edge weight between $i\in n$ and $j \in n$}
 \textbf{initialize} $\gamma=50/100$ and \emph{i}=0\;
 \For{each n}{
  \textbf{define} $l$ = integer(i/$\gamma$); $i$+=1\;
  \For{each n}{
      \If{n not labeled}{
       \textbf{compute} $t$ = neighbors of $n$\;
       \textbf{compute} $t_l$ = labeled neighbors of $n$\;
       \If{$|t_l| + l \geq t$}{
        \textbf{initialize} \textit{score}, $c$\\
        \For{each $t_i \in t$}{
            score += label $t_i * e_{n{t_i}}$; c += $e_{n{t_i}}$\\
            }
        \textbf{update} label $n = score/c$
       }
        }
    }
 }
\caption{Label Propagation Algorithm}
\label{algo:label_prop}
\end{algorithm}





\paragraph{} Once we have obtained hashtag-level or word-level polarity scores, we infer the polarity of a tweet or a group of tweets (e.g. all tweets by a given user) by averaging the polarity scores inferred by our algorithms for all the hashtags and words used in data subset. This approach is similar to the aggregation conducted in \cite{Holthoefer2015}, but our label propagation allows for the incorporation of thousands of words and hashtags, rather than relying on only a small hand-annotated set. If the data subset does not contain any of the keywords labelled by our algorithm (e.g. in a hashtag-based approach, the tweet contains no hashtags), we consider it unclassified. In some cases, primarily for evaluation, we convert the polarity scores into a ternary negative/neutral/positive position by using the cut-offs $\{<0, 0, >0\}$ for the $[-1, 1]$ scale and $\{<0.5, 0.5, >0.5\}$ for the $[0, 1]$ scale.

This methodology allows us to infer the polarity of any group of tweets along any dimensions, provided a small set of seed words or hashtags for each dimension. Thus, we can examine how polarities differed for different groups of users and how they changed over time. The two dimensions we focus on are Pro-India/Pro-Pakistan and Pro-Peace/Pro-Aggression. In practice we found that minor variations in the exact words in the seed set had no noticeable impact on our final results. For the network-based methods, we label Pro-India seeds as $+1$, Pro-Pakistan seeds as $-1$, Pro-Peace seeds as $+1$, and Pro-Aggression seeds as $-1$. For the embedding-based approach, we label Pro-India seeds as $+1$ Pro-Pakistan seeds as $0$, Pro-Peace seeds as $+1$, and Pro-Aggression seeds as $0$. For all word-based approaches, we limit the vocabulary size to 50K.\footnote{We provide our manually defined seed sets and label propagation code on \url{https://github.com/amantyag/india_pakistan_polarization/}.}

\begin{table}[!htb]
\centering
\caption {Classification results for the 100 most followed Indian and Pakistani Twitter accounts, where Pro-India or Pro-Pakistan are treated as the dominant class, and the nationality of the account owner is treated as a gold label. \%Unk denotes accounts that our algorithm was unable to classify and \%Incorrect denotes accounts that received polar opposite labels (e.g. Indian accounts classified as Pro-Pakistan)}
\begin{tabular}{lccccccccc} 

& \multicolumn{4}{c}{Pro-India (84 accounts)} & & \multicolumn{4}{c}{Pro-Pakistan (85 accounts)} \\
\cmidrule{2-5} \cmidrule{7-10}
 & Prec. & Recall & \%Unk.  & \%Incorrect & &  Prec. & Recall & \%Unk.  & \%Incorrect\\
\hline
\hline
Hashtag & 0.91	& 0.25 & 0.68 & 0.07 &	& 0.90 & 0.61 & 0.36 & 0.02 \\
Word & 0.69 & 0.69 & 0.24 & 0.07 & & 0.83 & 0.35 & 0.34 & 0.31\\
Sentprop & 0.48 & 0.80 & - & 0.20 & & 0.43 & 0.15 & - & 0.85 \\

\hline
\end{tabular}
\label{tab:tableeval}

\end{table}




\section{Evaluation}
\label{sec:eval}


\paragraph{\textbf{Automated Evaluation}} We first evaluate our methods by focusing on the Pro-India and Pro-Pakistan dimension and assuming that popular users in India are more likely to post Pro-India content and popular users in Pakistan are more likely to post Pro-Pakistan content. From the \url{Socialbakers.com} platform, we identified the 100 most followed Twitter accounts in India and in Pakistan.
16 of the Indian accounts and 15 of the Pakistani accounts do not occur in our data, leaving 84 Indian accounts with 2,199 tweets and 85 Pakistani accounts with 1,456 tweets for evaluation. For each account, we average word and hashtag polarities over all tweets from the account, and binarize the resulting score into a Pro-Pakistan or Pro-India position.

\Tref{tab:tableeval} reports results. Both of the network-based approaches rely on hashtag or word co-occurrences to propagate labels. Thus, hashtags and words that do not have any co-occurrence links to the original seed list are unable to be labeled. For instance, in the hashtag propagation approach, our method labels 41,700 hashtags out of 67,059 total hashtags in the dataset. Any users who only use unlabeled words or hashtags are therefore unable to be classified by our algorithm, resulting in 88/169 unlabeled accounts for the hashtag approach and and 49/169 unlabeled accounts for the word approach (\%Unk in \Tref{tab:tableeval}). In contrast, SentProp obtains polarity scores for all accounts, as it relies on embedding similarity and can propagate labels between words, even if they do not ever co-occur.

However, although SentProp labels more accounts, its precision is much lower than the network-based methods.
The network-based hashtag propagation approach overall obtains the highest precision and the least explicit errors -- lower recall scores occur because of accounts that it leaves unlabeled, rather than because of accounts that it labels incorrectly. Although the word-propagation approach labels more accounts and works well over the Indian accounts, its classification of the Pakistani accounts is close to random. We suspect that our method works well for hashtags, because they tend to be strongly polar and indicative of the overall sentiment of the tweet.
A word-based approach likely requires more careful handling of subtle language cues like negation or sarcasm.

In our subsequent analysis, we use the network-based hashtag propagation method in order to infer polarities, thus favoring high precision and strong polarization, and choosing not to analyze data where we cannot infer polarity with high-confidence. Additionally, in examining the data set, we found that many of the top-followed accounts in India and Pakistan consisted of celebrities who avoided taking stances on politicized issues, which makes the high number of unclassified accounts in this subset of the data unsurprising.


\begin{table}
    \centering
    \caption{Inter-annotator agreement and classification accuracy over 100 manually annotated data points}
    \begin{tabular}{c@{\hskip 0.2in}c@{\hskip 0.2in}c@{\hskip 0.2in}c@{\hskip 0.2in}c}
    \hline
        & Krippendorff $\alpha$ & \% Agree. & Hashtag Acc. & Soft Hashtag Acc. \\
        \hline
        \hline
        India/Pakistan & 0.77 & 88\% & 74\% & 89\% \\
        Aggression/Peace &  0.60 & 74\% & 57\% & 76\% \\
        \hline
    \end{tabular}
    \label{tab:manual_eval}
\end{table}

\paragraph{\textbf{Manual Evaluation}} In order to further evaluate our methods, we compare the performance of the network-based hashtag model with a small sample of manually annotated tweets. We randomly sampled 100 users from our data set. For each user, we randomly sampled 1 day on which the user tweeted and aggregated all tweets from that day. Thus, we conduct this evaluation at a per-user-per-day level. Two annotators  independently annotated each data sample as Pro-India/Pro-Pakistan/Neutral/Can't Determine and Pro-Peace/Pro-Aggression/Neutral/Can't Determine. For simplicity, we collapsed Neutral/Can't Determine and Unclassified into a single ``Neutral'' label. Notably, the Pro-Peace/Pro-Aggression and Pro-India/Pro-Pakistan dimensions are distinct. For example, users may write tweets that are Pro-Peace and Pro-Pakistan: \textit{``Don’t let people pull you into their War, pull them into your Peace... Peace for our World, Peace for our Children, Peace for our Future !! \#PakSymbolOfPeace \#SayNoToWar''} or that are Pro-Peace and Pro-India: \textit{``Very mature conciliatory speech by \#ImranKhan. We now urge him to walk the talk. Please return our \#Abhinandan safely back to us. This will go a long way in correcting perceptions and restoring peace. \#SayNoToWar''}.

\Tref{tab:manual_eval} reports inter-annotator agreement, which is generally high. Additionally, most disagreements occurred when one annotator labeled Neutral/Can't Determine and the other did not, meaning polar opposite annotations were rare. If we only count polar opposite labels as disagreements, the percent agreement rises to 94\% for both dimensions.

Then, the two annotators discussed any data points for which they initially disagreed and decided on a single gold label for each data point. We compare  performance of the network-based hashtag propagation method against these gold annotations in \Tref{tab:manual_eval}. In this 3-way classification task, the accuracy of random guessing would be 33\%, which our method easily outperforms. In particular, the ``Soft'' accuracy, in which we only consider the model output to be incorrect if it predicted the polar-opposite label, meaning neutral/unclassified predictions are not considered incorrect, is high for both dimensions.\footnote{We provide the manual annotations as well as additional metrics on \url{https://github.com/amantyag/india_pakistan_polarization}.}

\section{Results and Analysis}
\label{sec:results}
We investigate multiple aspects of our data set, including network structure, polarities of various entities, and changes over time. Based on prior work suggesting that political entities in India and Pakistan may use social media to influence public opinion \cite{indian_2014_elections,india_twitter_rahul,Kumar2016PoliticalMarket,pakistan_social}, we pay particular attention to the Twitter accounts of politicians as a method for uncovering political agendas.

\begin{table}[!htb]
\centering
\caption{Overall polarities of users and tweets.}
\begin{tabular}{lcc@{\hskip 0.1in}|@{\hskip 0.1in}lcc} 
\hline 
Position & Unique Users & Total Tweets & Position & Unique Users & Total Tweets \\
\hline
\hline
Pro-India & 125K (23\%) & 1.16M (46\%) & Pro-Aggression & 78K (14\%) & 626K (25\%)\\
Pro-Pakistan & 117K (20\%) & 764K (30\%) & Pro-Peace & 252K (45\%) & 1.48M (59\%)\\
Unclassified  & 325K (57\%) & 578K (23\%) & Unclassified & 237K (40\%) & 351K (16\%) \\
\hline

\end{tabular}
\label{tab:res_overal_pol}

\end{table}

\begin{figure*}
    \centering
    \includegraphics[width=0.30\linewidth]{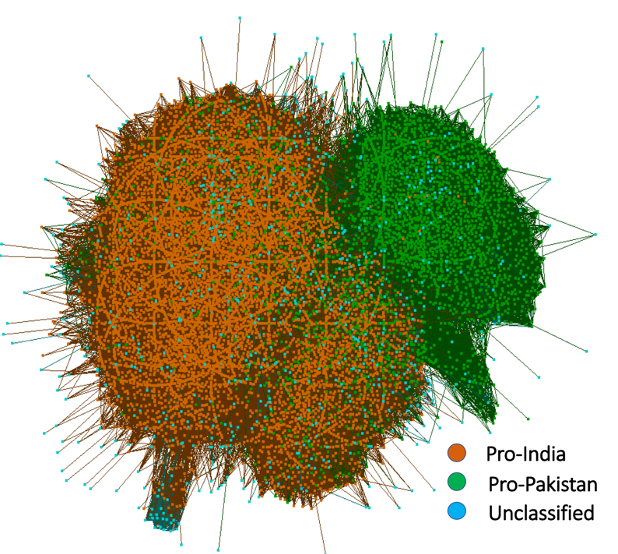}
    \includegraphics[width=0.30\linewidth]{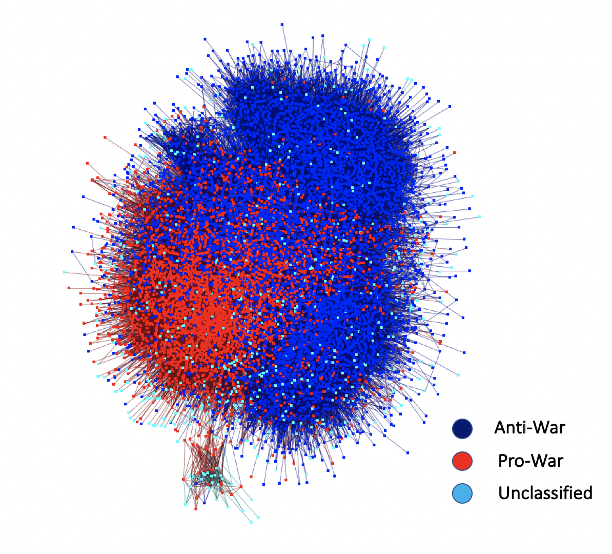}
    \caption{30-core all communication networks, colored by Pro-India/Pro-Pakistan polarity (left) and Pro-Peace/Pro-Aggression polarity (right). The Pro-India/Pro-Pakistan network displays more homophily than the Pro-Peace/Pro-Aggression network.}
    \label{fig:retweet_polarity}
\end{figure*}

\paragraph{What are the overall polarities of our data set?}
In \Tref{tab:res_overal_pol}, we obtain polarity scores for each user and tweet and then ternarize them into Pro-India/Pro-Pakistan/Unclassified and Pro-Peace/Pro-Aggression/Unclassified as in \Sref{sec:eval}. At the user level, the classified accounts are approximately balanced between Pro-India and Pro-Pakistan. However, at the tweet level, the classified data contains a high percentage of Pro-India tweets, suggesting Pro-India users tweeted about this issue more prolifically. Further, there is a much higher percentage of Pro-Peace users than Pro-Aggression users. This pattern also holds at the tweet level, where only a small percentage of tweets are unclassified.

\paragraph{What are characteristics of the communication network?}
Next, we examine the communication network between users, particularly prevalence of echo chambers. Did users with opposite positions interact? \Fref{fig:retweet_polarity} shows a 30-core all communication network constructed using ORA-PRO \cite{ora_tool_kit}. Accounts are colored based on their Pro-India/Pro-Pakistan polarity (left) and Pro-Peace/Pro-Aggression polarity (right). An edge occurs between two users if one user retweeted, mentioned, or replied to the other and users with $\le30$ links are not shown. Unsurprisingly, the Pro-India/Pro-Pakistan position is highly segregated, with little interaction between users with different positions. In contrast, the Pro-Peace/Pro-Aggression dimension is more mixed. Although there are some areas of high density for each position, there are interactions between users of different positions, which are potential avenues for users to influence each other's views.

\begin{figure}
    \centering
    \includegraphics[width=0.35\linewidth]{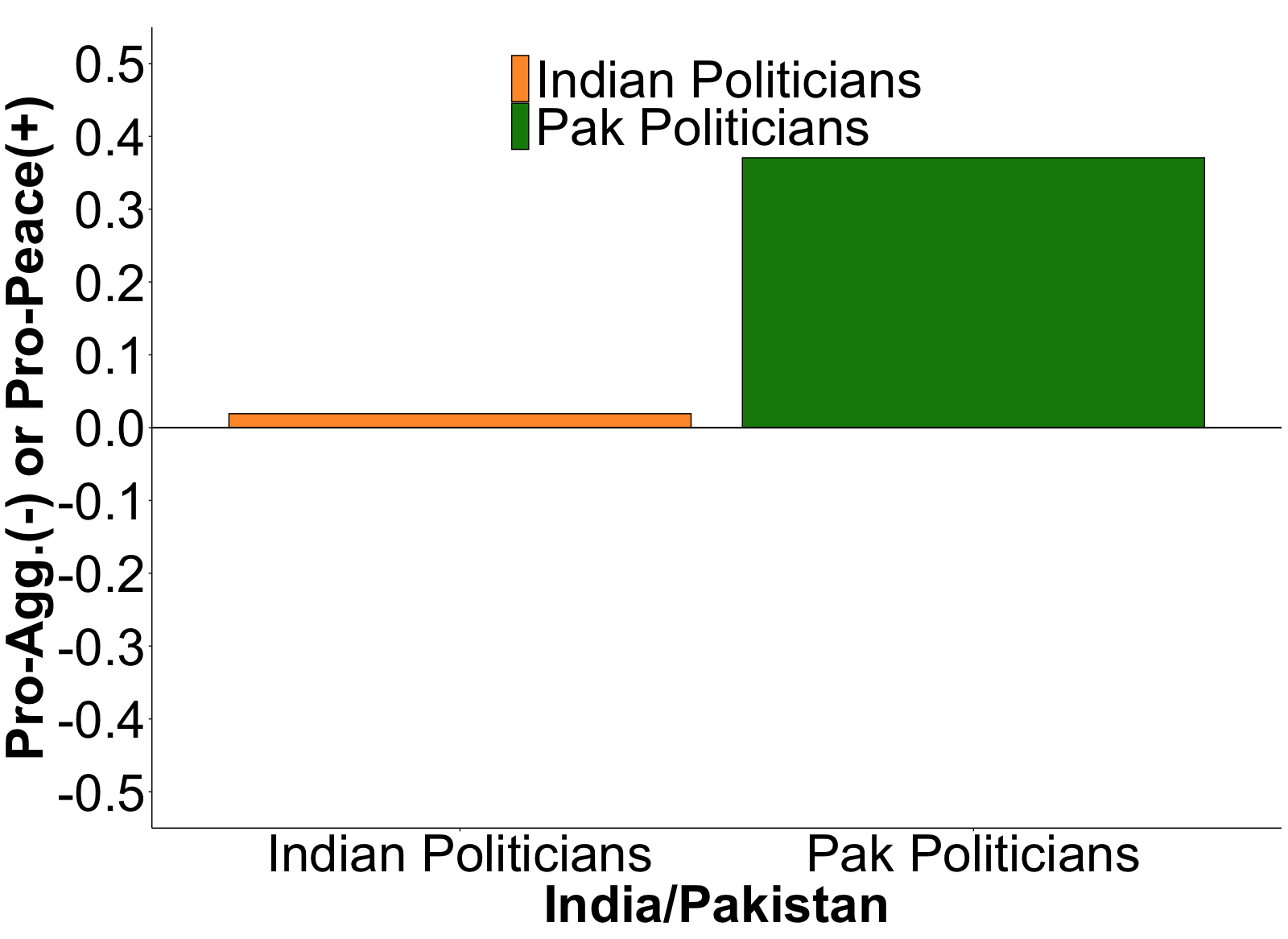}
    \includegraphics[width=0.35\linewidth]{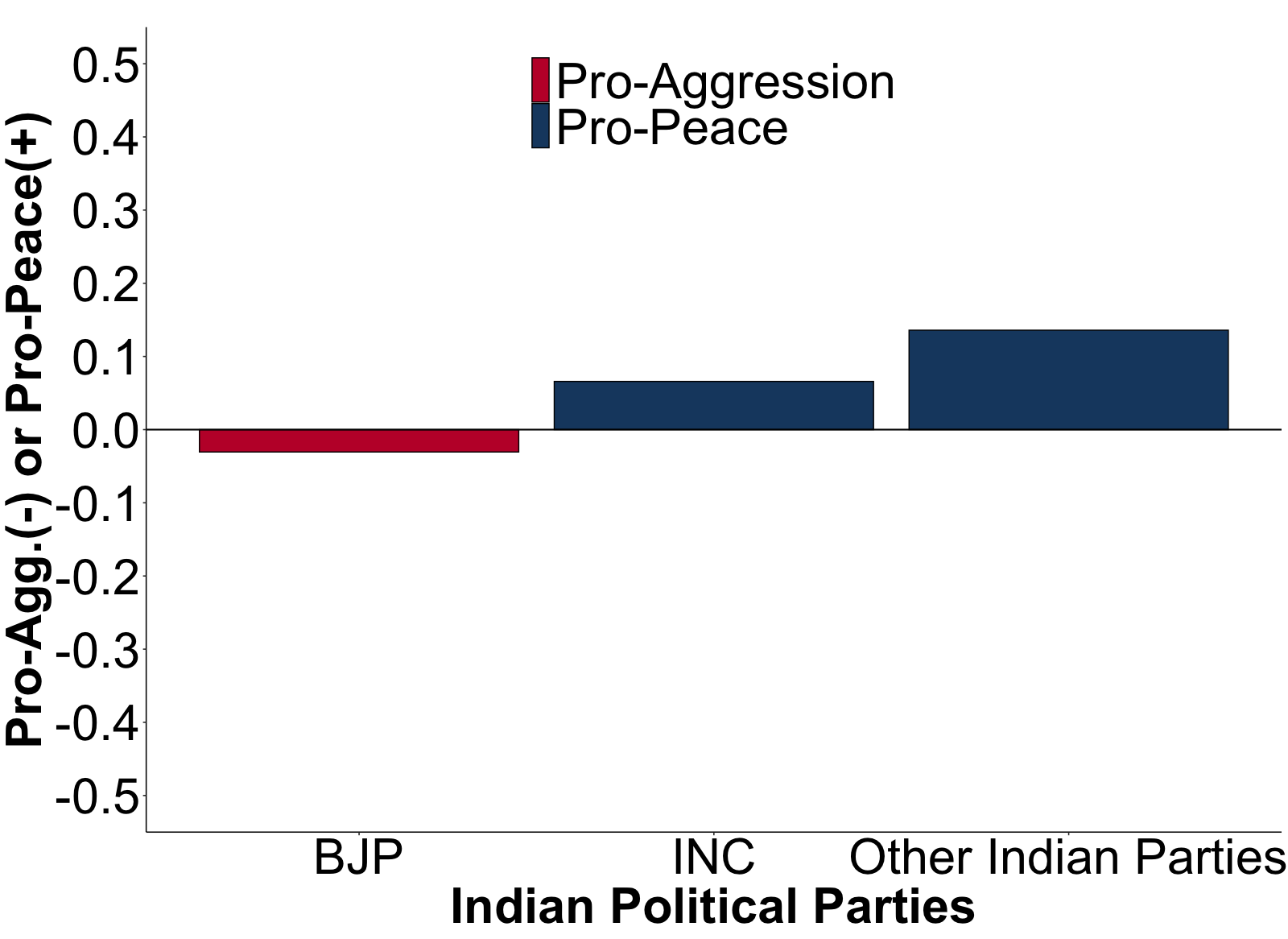}

    \caption{Aggregate Pro-Peace and Pro-Aggression polarities of the most popular Indian (33/78) and Pakistani (36/66) politicians in our data set (left) and of members of Indian political parties (right).}
    \label{fig:india_pak_polarity}
\end{figure}


\paragraph{How polarized were different political entities?}
We investigate the polarities projected by different political entities: specifically BJP politicians (currently in power in India), INC politicians (largest opposition party), other Indian politicians, and Pakistani politicians. We used the \url{Socialbakers.com} platform to obtain the Twitter handles of the 100 most followed politicians in India and Pakistan. Our data contained tweets from 66 Pakistani and 78 Indian politicians, and our hashtag model inferred scores for 36 Pakistani and 33 Indian politicians. \Fref{fig:india_pak_polarity} (left) reports aggregate polarity scores over all tweets from these politicians. Pakistani politicians were predominantly Pro-Peace, while Indian politicians expressed mixed polarities, yielding a near neutral score.

We then examined a broader set of Indian politicians, subdivided by political party based on a list of members running for parliament elections in 2019 \cite{loksabhaelections2019}. Out of the 1,360 Twitter handles in the list, our data set contained activity from 316 BJP accounts, 281 INC accounts, 204 other Indian party accounts.

\Fref{fig:india_pak_polarity} (right) shows the overall polarities, aggregated from all tweets by verified members of each party. Strikingly, members of the BJP party are positioned as much more Pro-Aggression than the members of either the INC or other parties, and the party overall obtains a Pro-Aggression polarity score. This score is not dominated by 1-2 strongly polarized members of the party: if we aggregate the polarity scores by individuals instead of by party, 15\% of BJP members had net Pro-Aggression scores and 13\% had net Pro-Peace scores, in comparison to 10\% Pro-Aggression/25\% Pro-Peace for INC, and 6\% Pro-Aggression/29\% Pro-Peace for other parties. The language used by BJP politicians was often openly Pro-Aggression: \textit{\#IndiaWantsRevenge We need to give a befitting reply to Pakistan, we will strike back...}.

These results support observations made by journalists and community members about the role of the BJP party in these events. BJP is well-known for promoting nationalism, and several journalists have speculated that conflict with Pakistan would increase Prime Minister Modi's chances of winning the upcoming elections in April and have accused the BJP of war-mongering \cite{ForbesModiElection,orfOpinion,theWeekWarMonger}.

\begin{figure*}
    \centering
    \includegraphics[width=0.85\linewidth]{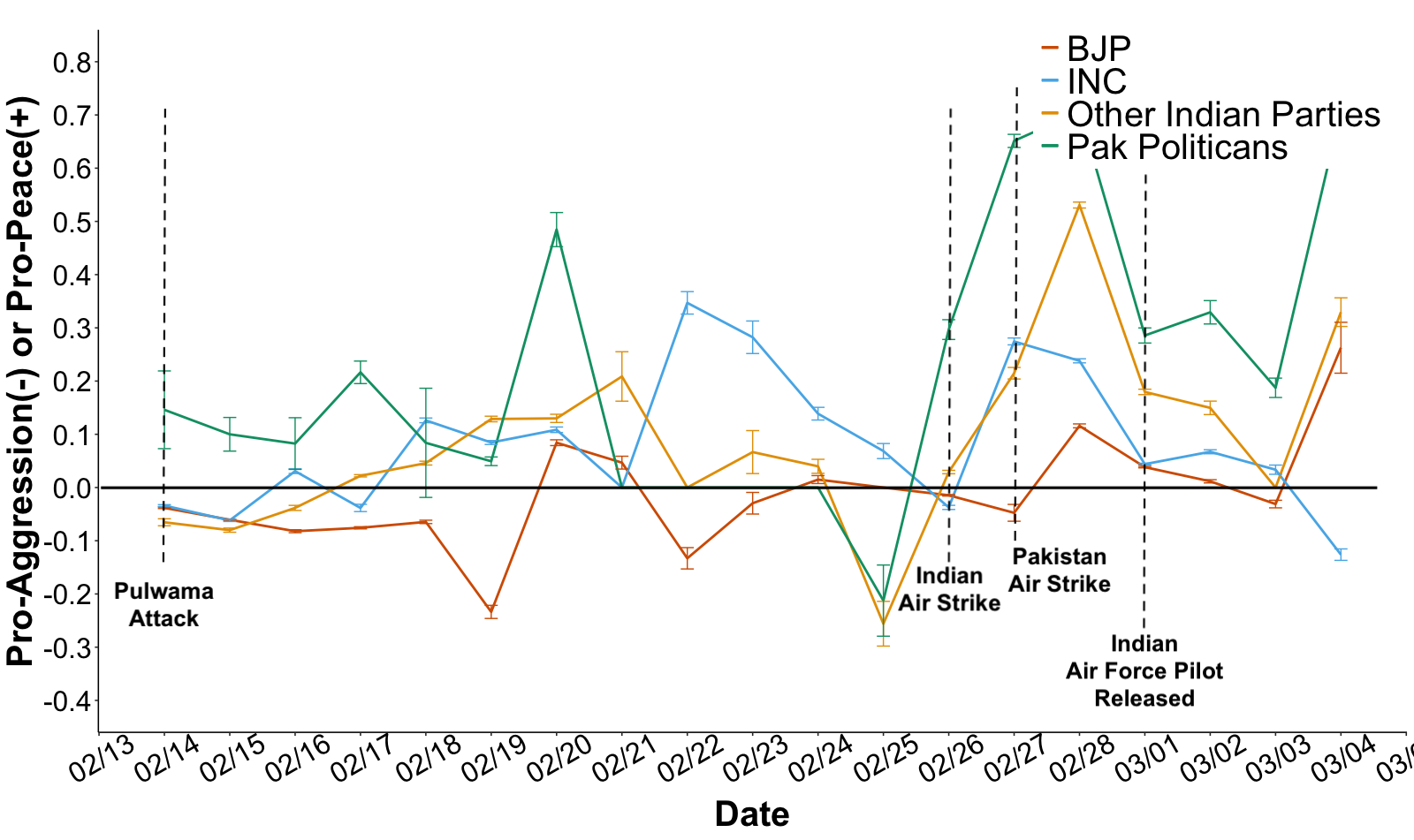}
    \caption{Daily Pro-Peace/Pro-Aggression positions of political entities. Negative values denote net Pro-Aggression polarity and positive values denote net Pro-Peace. The error bars represent ±1 standard deviation.}
    \label{fig:polarity_time}
\end{figure*}

\paragraph{How did polarization change over time?}
\Fref{fig:polarity_time} shows how this polarity changed over the two-week period of events: we infer a Pro-Peace/Pro-Aggression polarity score for all tweets posted by members of the specified political subgroup, and we plot the average score across tweets posted each day. 

Immediately following the initial attack on 2/14, the tweets from all Indian political party members are inclined towards Pro-Aggression, suggesting initial outrage. However, over the next few days, while tweets from INC and other Indian political party members switch towards Pro-Peace, tweets from BJP politicians remain consistently Pro-Aggression. There is high volatility between 2/20 and 2/26. However, there was a much lower volume of tweets about the Pulwama incidents during this time period,\footnote{Tweet volume is provided in our \href{https://github.com/amantyag/india_pakistan_polarization/}{project repository}.} and we do not believe these fluctuations are meaningful. The volume of tweets increases once again following the Indian (2/26) and Pakistani (2/27) airstrikes. Tweets by Pakistani politicians generally fall on the Pro-Peace side, but they become more polarized after the Indian airstrike and reach a peak following the Pakistani airstrike. This is consistent with reported quotes by Pakistani officials (\Sref{sec:events}), saying that the airstrike was designed to avoid escalation. Similarly, tweets by Indian politicians from the INC and other parties become strongly Pro-Peace directly following the Indian airstrike, with polarity increasing after the Pakistani airstrike. In contrast, on the day of the Pakistani airstrike, tweets by BJP politicians remain Pro-Aggression, possibly focusing either on praise for the Indian airstrike or condemnation of the Pakistani airstrike. The polarity of the BJP tweets belatedly switches to Pro-Peace on the following day (2/28), though the strength of the Pro-Peace polarity still remains weaker for BJP tweets than for tweets by other politicians.

\section{Discussion and Related Work}

The potential that social media platforms have for manipulating public opinion has led to growing interest in information operations and the development of social cyber security as a field of research \cite{carley2018social,starbird2019disinformation}. While we do not claim that social media coverage of the Pulwama incident constituted an information operation, e.g. coordinated efforts to manipulate public opinion and change how people perceive events \cite{starbird2019disinformation}, we do find similarities between our observations and other work in this area. Notably, as described in \Sref{sec:events}, the Indian and Pakistani governments maintain starkly different accounts about the events that occurred, particularly whether or not the 2/26 airstrikes resulted in 200 casualties. Similarly, Russian and Ukranian governments circulated conflicting narratives about the cause of the crash of Malaysian Airlines Flight MH17 in 2014, which prompted analyses of information operations about this incident. In a work similar to ours, \cite{golovchenko2018state} examine social media coverage of the incident by using a set of hashtags to collect all relevant tweets during a set time frame. Other work has examined the media influence of Chinese and Russian state actors in various domains, including US and UK elections and the Syrian War \cite{field2018framing,king2017chinese,kriel2019reverse,rozenas2019autocrats,starbird2018ecosystem}. \cite{arif2018acting} examine Russian influence in polarizing movements on Twitter, particularly the \#BlackLivesMatterMovement, and observe how Russian actors attempted to increase tensions between involved parties. Furthermore, the polarization that we observe in our data align with the ``Excite'' and ``Dismay'' strategies, which are tools of public opinion manipulation described in the BEND forms of maneuver \cite{beskow2019social}.


Almost all of these works are focused on U.S. social media, possibly involving Chinese and Russian actors. In general, most work on polarization and public opinion change has focused on U.S. politics \cite{chen2019seeing,darwish2019quantifying,khosla2019events}, with a few exceptions focusing on Germany \cite{darius2019twitter}, Egypt \cite{Holthoefer2015,Weber2013}, and Venezuela \cite{morales2015measuring}. Work on social media in India and Pakistan has focused on healthcare \cite{Abbasi2018}, natural disasters \cite{Murthy2013}, self-promotion (e.g. ``brand marketing'') primarily in relation to elections \cite{indian_2014_elections,india_twitter_rahul,Kumar2016PoliticalMarket}, or on election forecasting \cite{Kagan2015,pakistan_social}, though \cite{pakistan_army_twitter} does argue that the Pakistan Army uses social media to subvert democracy. While these works only focus on intra-country analysis, our work also examines tensions between India and Pakistan. A small selection of work has also looked at the incidents in Pulwama and the implications of rising tensions. \cite{feyyaz2019contextualizing}  and \cite{pandya2019future} discuss the sociopolitical context and implications of events from a non-computational perspective. \cite{palakodety2019kashmir} additionally conduct a social media analysis, but they use YouTube data and focus on identifying deescalating language. Their timeline of escalation and deescalation is generally consistent with our findings.

Our primary methodology involves using label propagation to infer aggregated polarity scores. 
In language corpora, label propagation has typically relied on embedding similarity \cite{hamilton2016inducing,rothe2016ultradense}. Instead, our approach takes advantage of the short-text nature of Twitter through co-occurrences networks, as well as the strong semantic signals provided by hashtags \cite{ICWSM1510473}. Prior methods for analyzing polarization focus on inferring user-level scores \cite{darius2019twitter,darwish2019quantifying} or require in-language annotations and feature-crafting \cite{magdy2016isisisnotislam}, whereas our method facilitates analyzing how user polarities can change over time in a multilingual corpus.




\textbf{Conclusions}
Polarizing language on social media can have long-lasting sociopolitical impacts.
Our analysis shows how Twitter users in India and Pakistan used polarizing language during a period of escalating tensions between the two nations, and our methodology offers tools for future work in this area.

\section*{Acknowledgements} We thank anonymous reviews and colleagues who provided feedback on this work. The authors would like to acknowledge the support of center for Computational Analysis of Social and Organizational Systems (CASOS), Carnegie Mellon University. This research was also supported in part by Public Interest Technology University Network Grant No.~NVF-PITU-Carnegie Mellon University-Subgrant-009246-2019-10-01. The second author of this work is supported by NSF-GRFP under Grant No.~DGE1745016. Any opinions, findings, and conclusions or recommendations expressed in this material are those of the authors and do not necessarily reflect the views of the NSF.


\bibliographystyle{splncs04}
\bibliography{references.bib}

\end{document}